\documentclass{styles/svproc}
\usepackage{url}

\usepackage{cite}
\usepackage{comment}
\usepackage{amsmath,amssymb,amsfonts}
\usepackage{algorithmic}
\usepackage{graphicx}
\usepackage{textcomp}
\usepackage{xcolor}
\usepackage{booktabs}
\usepackage{lipsum} 
\usepackage{url}
\usepackage{hyperref}
\usepackage[disable]{todonotes}
\def\BibTeX{{\rm B\kern-.05em{\sc i\kern-.025em b}\kern-.08em
    T\kern-.1667em\lower.7ex\hbox{E}\kern-.125emX}}

\begin{document}
\mainmatter              % start of a contribution
\title{Towards automated open source assessment - An empirical study}
\titlerunning{Automated OSS assessment}  % abbreviated title (for running head)
%                                     also used for the TOC unless
%                                     \toctitle is used
%
\author{Sai Pranav Koyyada\inst{1} \and Denim Deshmukh\inst{1}
Deepika Badampudi\inst{1} \and Vida Ahmadi\inst{1},\inst{2} \and Muhammad Usman\inst{1} }
\authorrunning{Sai Pranav et al.} % abbreviated author list (for running head)
%
%%%% list of authors for the TOC (use if author list has to be modified)
%\tocauthor{Ivar Ekeland, Roger Temam, Jeffrey Dean, David Grove,
%sCraig Chambers, Kim B. Bruce, and Elisa Bertino}
%
\institute{Blekinge Institute of Technology, Sweden\\
\email{saky19@student.bth.se, dede19@student.bth.se, deepika.badampudi@bth.se, vida.ahmadi.mehri@bth.se, muhammad.usman@bth.se},\\ %WWW home page:
%\texttt{http://users/\homedir iekeland/web/welcome.html}
\and
City Network International AB, Sweden}

\maketitle              % typeset the title of the contribution

\begin{abstract}
The open source software (OSS) assessment has become important given the increased adoption of OSS in commercial product development. Researchers proposed many OSS assessment models. However, little is known about the industrial relevance of the models. In this study, we proposed an automated tool based on the OSS assessment attributes identified together with a European cloud provider company. We analyzed 51 repositories to observe patterns in maintenance activities over their lifetime (from inception to the latest release). Based on the analysis, we propose a novel approach for evaluating the maturity of the OSS project. Finally, we assessed the usefulness of our automated solution in a pilot study.
\end{abstract}
\section{Introduction}
Software companies increasingly adopt Open Source Software (OSS), which has become part of the mainstream practice in software engineering \cite{8880574}. However, the selection of OSS is still challenging \cite{lenarduzzi2020open}. The practitioners in our case company: a European cloud provider, reported similar challenges. They mentioned gathering information from multiple sources and tools as a complex and time-consuming activity. The case company identified the need to automate the OSS assessment to reduce the selection effort. \\
The automation of the OSS assessment requires the identification of the attributes practitioners consider for OSS selection. 
%It is important to identify the attributes considered for OSS selection for automation.
In the last two decades, many OSS quality models have been proposed to  assist the OSS selection. Lenarduzzi et al. \cite{lenarduzzi2020open} identified discrepancies in the information provided in the evaluation models and the practitioners' information needs. In addition, little is known about how relevant these models are in practice as they have not been validated extensively\cite{lenarduzzi2020open}. The OSS quality assessment models suggest many evaluation attributes. Maintenance is one of the most considered quality attributes in the OSS assessment models proposed in the previous studies \cite{yilmaz2022quality}. Li et al. \cite{li2022exploring} conducted a survey to understand the attributes practitioners consider important in OSS selection. Practitioners mention maintenance as an important attribute; however, they did not mention metrics for assessing maintenance \cite{li2022exploring}. Metrics are important to automate quality assessment. However,  practitioners did mention metrics for assessing software maturity, such as the number of forks, number of releases, and number of commits \cite{li2022exploring}. However, Li et al. \cite{li2022exploring} suggested that while some practitioners consider the number of commits as a metric to evaluate software maturity, evaluating the prevalence of commits over time and the types of commits may be more useful.\\
Levin and Yehudai \cite{levin2017boosting} proposed a model to classify the comments based on the maintenance activities. However, their motivation to classify was to improve planning and resource allocation for maintenance. As indicated by Li et al. \cite{li2022exploring}, the prevalence of commits over time and types of commits could be a good measure of maturity. However, they did not find any portals that effectively provide community-related factors to automate OSS project assessment \cite{li2022exploring}. Therefore, it is interesting to investigate how commit classification based on maintenance activities can help automate OSS assessment. \\\\
The study aims to identify commonly considered attributes in OSS selection in the case company and investigate to what extent commit classification based on maintenance activities can help in the OSS assessment. Many models for commit classification exist \cite{hindle2009automatic, gharbi2019classification, levin2017boosting, ghadhab2021augmenting}. Our objective is not to propose the most accurate classification model but to demonstrate the use of commit classification in the OSS assessment. We used Levin’s commit classification model \cite{levin2017boosting} to monitor different maintenance activities carried out in OSS projects. We analyzed 51 OSS projects, including frameworks, APIs, libraries, databases, and applications, to collect attributes that can help facilitate the OSS assessment. Finally, we conducted a pilot qualitative study to understand practitioners’ opinions on the usefulness of the commit classification based on maintenance activities and commonly considered selection attributes in the OSS assessment.
\begin{comment}

\section{Related work}
Prior OSS assessment models proposed in the literature focus on assessing attributes such as license obligation, security, and code quality \cite{lenarduzzi2020open}. Community-related metrics such as the number of forks and the average time to release can provide information regarding the growth of the OSS project community \cite{li2022exploring}. However, no portals effectively provide community-related factors to automate OSS project assessment \cite{li2022exploring}. Li et al. \cite{li2022exploring} suggest considering the type of commits instead of the number of commits to get more meaningful insight into the OSS project maturity. Researchers have classified the type of commits based on maintenance activities \cite{hindle2009automatic, gharbi2019classification, levin2017boosting, ghadhab2021augmenting}. However, to the best of our knowledge, the use of commit classification for assessing OSS projects has not been validated so far.  
\end{comment}
\section{Research methodology}
Our goal is to answer the following research questions - 
\begin{enumerate}
    \item[RQ1:]What OSS attributes are considered important to automate in the case company?
    \item[RQ2:]How can commit classification be used to facilitate OSS assessment?
    \item[RQ3:]How do practitioners perceive the usefulness of our automated solution?
\end{enumerate}
We used design science method\cite{wieringa2014design} to answer the above research questions. 
\begin{figure}
    \centering
    \includegraphics[scale=0.7]{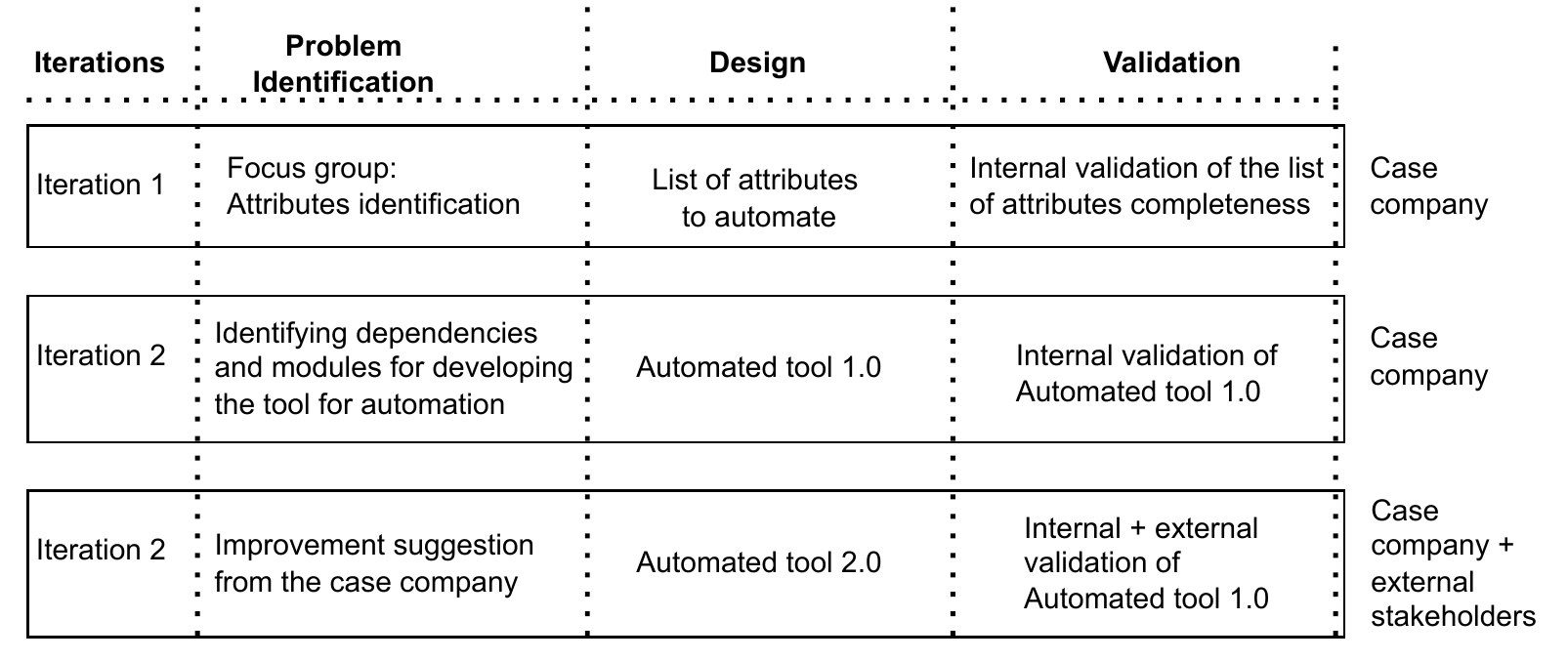}
    \caption{Research design followed in the study}
    \label{fig:RM}
\end{figure}
Figure \ref{fig:RM} depicts the iterations in the solution development and validation. The steps carried out in each iteration are described as follows . 
\begin{itemize}
    \item \textbf{Iteration 1:} In this iteration our goal is to identify the attributes that the case company considers important for automating OSS assessments. We used focus groups where key stakeholders from the case company and the authors discussed the different attributes. The input to the focus group was the case company's checklist for OSS assessment and the attributes frequently reported in the literature. The goal of the focus group was to identify attributes that can be automated to improve the efficiency and effectiveness of the OSS assessment. The outcome of the focus group was the list of attributes to automate. The first two authors were employed at the case company, providing them easy access to the developers involved in the OSS assessment. For internal validation, the first two authors discussed with three developers on the completeness of the attributes. Overall the developers agreed with the list of attributes.
    \item \textbf{Iteration 2:} We identified the different sources to retrieve the required attributes. We used various API Endpoints such as GitHub REST API, Stack Exchange REST API, python libraries such as pydriller, OWASP Dependency Checker, and other OSS solutions to gather the relevant information. We built an automated solution that could be triggered with a single command and return all results from the assessment in the JSON format which we refer to as Automated tool 1.0 in Figure \ref{fig:RM}. We discuss the tool in a focus group with the security expert and three developers. Our solution used the JSON files generated output to present the assessment results. The focus group participants from the case company requested a feature that pooled all the results on one page. 
    \item \textbf{Iteration 3:} The problem identification for this iteration was the input from the validation in Iteration 2. Therefore, we created a module that could show the results from various JSON files generated from our automated solution on one page which we refer to as Automated tool 2.0 in Figure \ref{fig:RM}. The source code, modules used in the automated solution, and the documentation of the tool usage are provided online\footnote{https://github.com/SaipranavK/oss-recon}{}. We verified the functioning of our automated tool by assessing 51 different OSS projects. Our automated solution generated the expected results. We presented the automated solution through a technical demo at the case company. During the technical demo, we assessed two OSS: one framework and one application. The results took under a minute to generate. We finally interviewed 10 developers from three different companies (including the case company) to validate the completeness of the assessment attributes and the usefulness of the automated tool.
\end{itemize}
\section{Results}
This section presents our results from the focus group study, analysis of commit classification, and preliminary validation based on practitioners' perceptions of our automated solution. 
\subsection{OSS assessment attributes}\label{Sec:OSS_Attributes}
Identifying the attributes required in the OSS assessment to automate the process is important. We discussed the attributes, the required metrics, and the information to automate the OSS assessment in the focus group. The input to the focus group was the collection of attributes and metrics frequently considered in the literature and the case company. We selected the attributes and metrics based on their importance and the ease of understanding perceived by the case company. The case company preferred using descriptive representation than numerical metrics-based representations. Therefore, the case company did not want information on traditional metrics like code complexity, coupling, cohesion, and other similar metrics. This section presents the attributes for assessing the OSS projects. 
\begin{table}[h]
\caption{OSS assessment attributes considered at the case company}
\label{tab:OSS_Assessment1}
\begin{tabular}{@{}p{5cm}p{7cm}@{}}
\toprule
\textbf{Attributes} & \textbf{Metrics and information}                                                                               \\ \midrule
Repository information     & Name, description, topics, API URL, programming languages, and Github community health percentage. \\
Repository activeness         & Age, last updated date, average time to release, number of open issues, active/recent releases, The commit activity, commit classification.             \\
Security                    & Vulnerabilities.                                                                                   \\
Community interest & Stars, forks and watchers. \\
Support                     & Stack overflow QAs.                                                                               \\
%Community interest          & Stars, forks, watchers.                                                                            \\
Legal requirements          & License type, permissions, conditions and limitations.                                             \\
%Maintenance                 & Commit activities and code frequency.                                                              \\

 \bottomrule
\end{tabular}
\end{table}
Table \ref{tab:OSS_Assessment1} contains the attributes considered important for assessing OSS projects by the case company. We also present the metrics and information gathered to access the attributes (see the second column in Table \ref{tab:OSS_Assessment1}). 

\textbf{Repository information:} The company starts assessing the OSS project by reviewing general repository information (see details on repository information in Table \ref{tab:OSS_Assessment1}).\\
\textbf{Repository activeness:} In addition to the generic information, the company reviews the repository's activeness by reviewing the average time it takes to release a new version, the number of open issues, and the list of active or recent releases of the OSS project. In addition to the company's attributes, we added age, last updated date, commit activities: the number of deletions and additions, and commit classification: the number of corrections, adaptions and perfective activities metrics. 

\textbf{Security:} The security expert at the case company identified security as an important criterion for adopting OSS projects. 

\textbf{Community interest:} In addition, the company reviews the support availability by considering the number of questions and answers posted with tags associated with the OSS project on StackOverflow.

\begin{figure}[!tbp]
  \centering
  {\includegraphics[width=\textwidth]{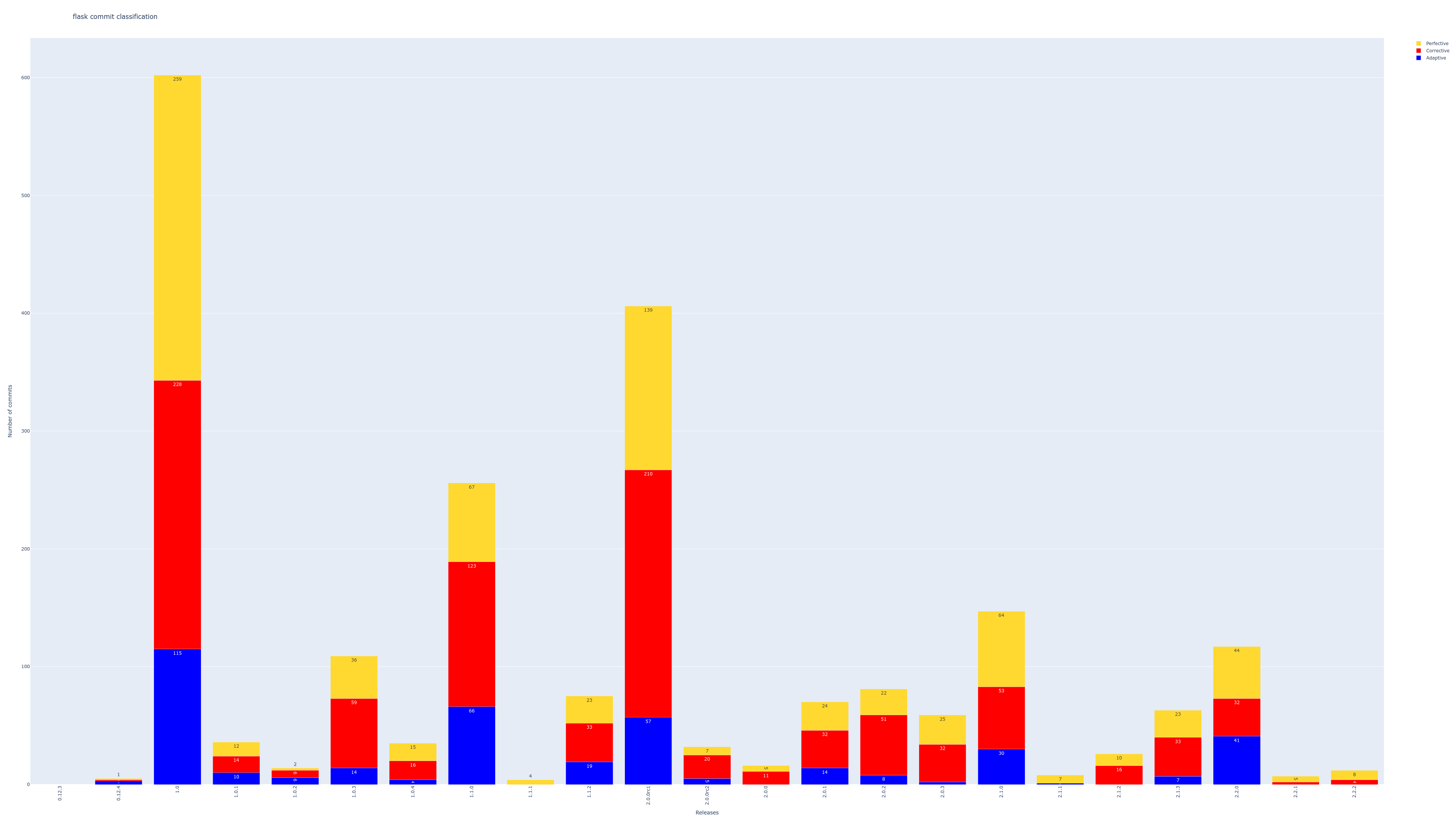}}
  \caption{Commit classification of OSS flask.}
  \label{fig:commitclassification}
\end{figure}
\begin{figure*}[!h]
  \centering
  {\includegraphics[scale=0.20]{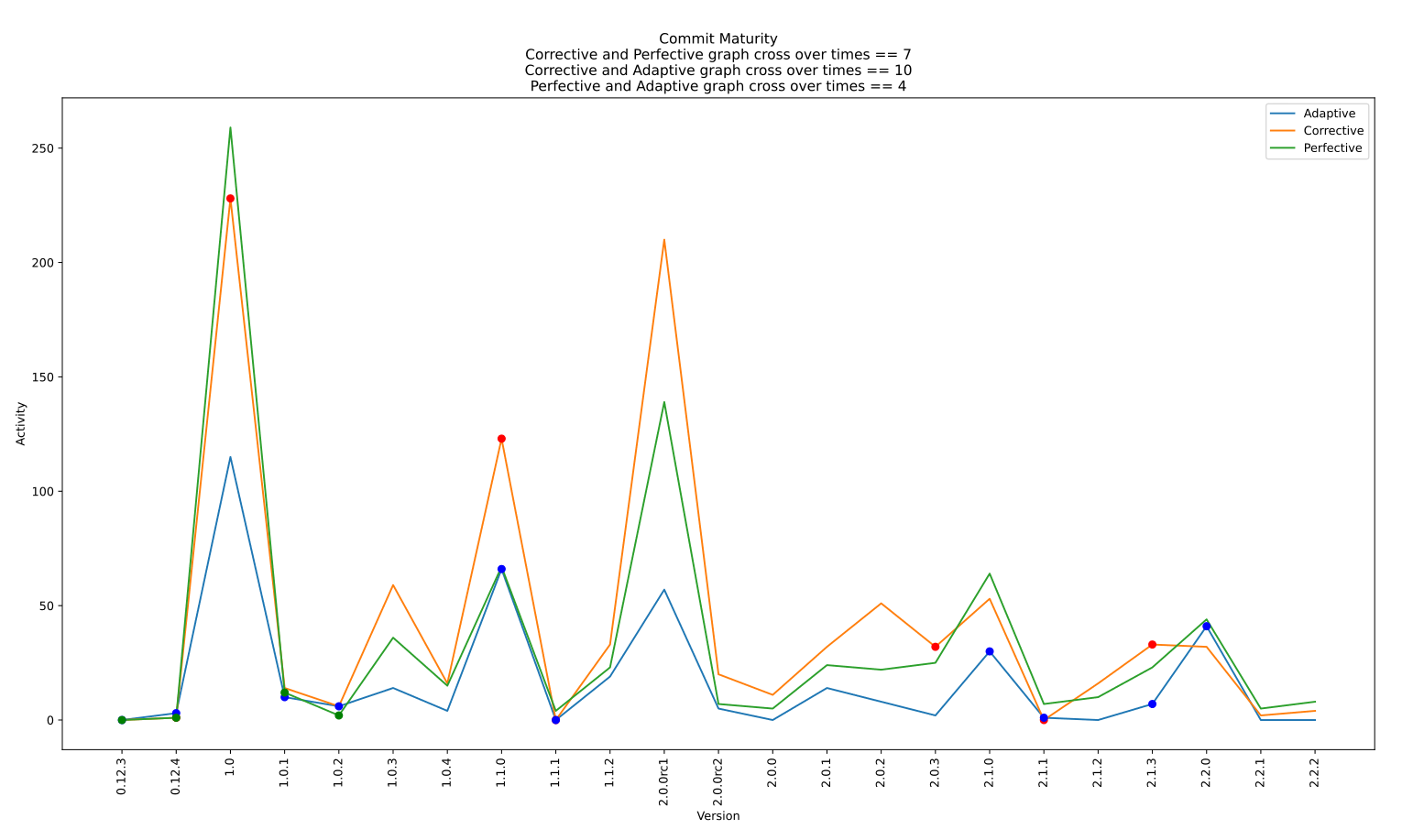}}
  \caption{Commit maturity of Flask repository.}
  \label{fig:CommitMaturityFlask}
\end{figure*}
\begin{figure*}[!tbp]
       % \centering
        {\includegraphics[scale=0.20]{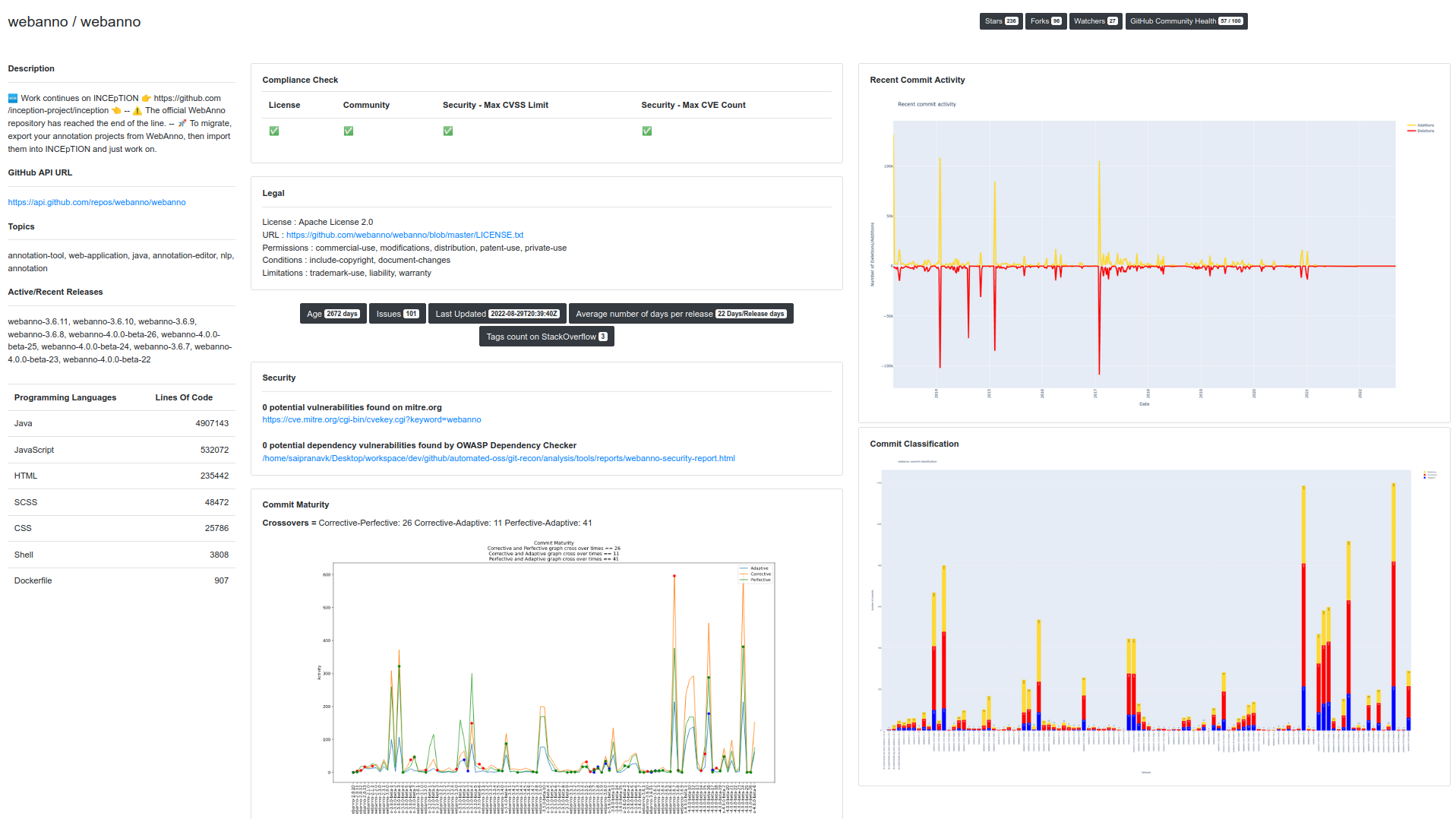}}
        \caption{Sample report from our automated solution for OSS webanno.}
        \label{fig:screenshot}
\end{figure*}
\subsection{Commit classification based on maintenance activities to evaluate OSS projects}
%\todo[inline]{Lehman's law}
We propose using commit classification, among other metrics, to evaluate OSS projects. Each new release of an OSS has additions or deletions compared to the previous version. These additions and deletions are changes that introduce new features, fix bugs, or extend the support of OSS. The changes implemented in a release are called maintenance activities.
The Software Engineering Body of Knowledge(SWEBOK) and IEEE14764 categorize software maintenance activities as corrective, adaptive, perfective, and preventative. Preventative and corrective activities are corrections, given their purpose to fix latent and operational faults. Perfective and adaptive activities are modifications to improve the software. 

We visualized the commit activeness based on the maintenance activities from the inception to the latest release of the OSS Project. We used the approach proposed by Levin et al.\cite{levin2017boosting} to classify the maintenance activities. Figure \ref{fig:commitclassification} shows the commit classification of an example OSS across its different releases. Each release has a distribution of three activities: corrective in red, perfective in yellow, and adaptive in blue. Preventative activities are proactive activities, and there are comparatively difficult to capture for classification. Therefore, our study did not include classification based on preventative activities. 

\subsubsection{Commit maturity: A novel perspective on commits classification}
We analyzed 51 OSS repositories, including frameworks, APIs, libraries, databases, and applications. The visualizations on commit classification on 51 open source repositories used for analysis are provided online\footnote{https://doi.org/10.5281/zenodo.7053198}. We observed the maintenance activities of the repositories from their inception to the latest release. The analyzed 51 OSS included very popular repositories with more than 10000 stars, popular repositories with over 5000 stars, and some growing OSS with over 500 stars on GitHub.

We observed adaptive activity is the least performed activity in each release for all the OSS repositories. Corrective and perfective activities have the most variance, i.e., the number of corrective activities may be higher than perfective in some commits and lower in others. When a certain maintenance activity's frequency goes down, and another maintenance activity's frequency goes up, we call it a crossover. 

We counted the number of crossovers for each combination of the maintenance activities and mapped them with the type of the OSS. We noticed that all OSS except the ones of type frameworks had a similar number of crossovers for each combination of maintenance activities. %Figure \ref{fig: Crossovers} shows a visualization of the maintenance activities crossovers for the 51 repositories.

Lehman's law \cite{lehman1980programs} suggests that a system should continuously change to remain useful. A change can be measured by the number of counts of corrective, adaptive, and perfective requests over a certain period \cite{barry2007software}. Any software project should have good features with minimum bugs and extensive support. Whenever a crossover happens, it indicates that the focus of the community shifts from one maintenance activity to another to make sustainable progress. If the community only focuses on adding new features, then there may be many bugs making the OSS unusable. The same applies when the community is only resolving bugs in the OSS. It means that the OSS has many bugs to be addressed and does not introduce new features to improve its value. The balance between the maintenance activities should be maintained. Frameworks are unique in this regard because of the scale of features and platforms they support. As we saw from our analysis, it may not be possible to maintain the balance between the maintenance activities for frameworks. We hypothesize that the total number of crossovers of each combination of maintenance activities should ideally be similar to the total number of releases. Crossover between Adaptive and Corrective activities =
\begin{equation}
 (A{i-1} > C{i-1}) \cdot (A{i}< C{i})
\end{equation}
Where (A{i}) is a list of Adaptive Activities and (C{i}) is a list of Corrective Activities for each version of the software. Similarly, we can calculate three types of intersections. 
The number of intersections between each pair of activities, i.e., Adaptive, Corrective, and Perfective, defines commit maturity.

Commit maturity is the number of times each maintenance activity crosses other maintenance activities over the project life cycle. It will allow practitioners to see if the OSS project maintains the balance between the maintenance activities. Figure \ref{fig:CommitMaturityFlask} shows the commit maturity and the number of crossovers for each crossover pair. Each dot represents an occurrence of a  crossover in a release. In this example, in release 1.0, the corrective activities decreased, and the perfective activities increased, resulting in a crossover. The flask project has a total crossover count of 21 out of 23 releases. Based on our hypothesis, it is a good maturity indication, and the popularity of the flask project is a testimony to it.

The case company requested all the information needed to evaluate the OSS repository on one page. Figure \ref{fig:screenshot} provides a screenshot of our solution. The left panel includes information on the repository, while the middle panel provides information on the metrics to assess security, support, and legal requirements. The middle panel also includes metrics to evaluate repository activeness: age, last updated date, the average time to release, and the number of issues. The commit activity, commit classification, and commit maturity are represented in graphical format. Finally, the community's interest: stars, forks, and watchers are presented in the top right corner.

\subsection{Practitioners perception of the automated OSS assessment and commit activeness}

\textbf{Completeness of the evaluation attributes:} we aimed to evaluate if the practitioners found our automated solution useful for automatically assessing OSS. All the participants agreed that the tool could help in the OSS assessments. Some of the participants wanted to see more attributes. For example, one of the interviewees mentioned \textit{"I want to gather more information like its compatibility with different operating systems and the tutorials sources."}. Since we designed our solution primarily for use in the case company, it is not surprising that interviewees wished for additional attributes. One solution could be to create a configurable solution where the stakeholders can select the important attributes in the assessment. \\
\textbf{Ease of understanding the information on OSS assessment attributes:} We explained the attributes to the interviewees, particularly the attributes such as commit maturity. We asked the interviewees if the attributes we used were easy to understand. All the participants unanimously agreed that our attributes were easy to understand. The interviewees added that \textit{"The attributes were easy to understand. It was simply like a GitHub page but with more information."}. In addition, the interviewees were positive about the new attributes such as commit maturity: \textit{"The commit evolution and maturity was something new but were still very easy to understand along with other attributes."}.\\
\textbf{Commit classification and maturity:} We asked the interviewees if commit classification and commit maturity are good visualizations to support the OSS adoption decision. All the interviewees agreed that the visualization was useful once we explained the attributes to them. One of the interviewees mentioned \textit{"Managers would love such a visualization because it not only is simple to understand but also will help a non-technical person easily comment on the OSS community."}.

\section{Conclusion and future work}

We reported initial findings from an empirical study to support the practitioners in the OSS assessment process. With the help of the practitioners in the case company, we first identified the attributes that could be automated in performing the OSS assessment. Our tool automatically collects and presents the data about the identified attributes in one place to facilitate the practitioners in performing the OSS assessment. We also investigated how commit classification based on different maintenance activities can be used in OSS assessment. We introduce the use of commit maturity to see if an OSS project is balanced in feature enhancements and bug fixes or overly focused on only one type of maintenance activity (e.g., only fixing bugs in case of corrective maintenance). We used our tool to analyze 51 OSS projects. We also shared the results with the company practitioners, who found our tool helpful in performing the OSS assessments. In the future, we plan to study commit maturity as a metric to assess OSS maintainability through extensive validation and application and standardize it for wider adoption.
Additionally, we will continue to enhance our automated OSS assessment tool by improving the range of supported attributes and desired metric outputs. We also wish to employ repository mining techniques to identify and correlate community activities with the OSS engagement and growth trends to comment on its popularity and support. Another interesting direction of research would be correlating maintenance activities for an OSS with its traditional maintainability metrics that can help evaluate the relationship between maintainability and maintenance activities, if any, and thus result in more branching paths of research in software metrics and the maintainability domain.

% Please number citations consecutively within brackets \cite{b1}. The 
% sentence punctuation follows the bracket \cite{b2}. Refer simply to the reference 
% number, as in \cite{b3}---do not use ``Ref. \cite{b3}'' or ``reference \cite{b3}'' except at 
% the beginning of a sentence: ``Reference \cite{b3} was the first $\ldots$''

% Number footnotes separately in superscripts. Place the actual footnote at 
% the bottom of the column in which it was cited. Do not put footnotes in the 
% abstract or reference list. Use letters for table footnotes.

% Unless there are six authors or more give all authors' names; do not use 
% ``et al.''. Papers that have not been published, even if they have been 
% submitted for publication, should be cited as ``unpublished'' \cite{b4}. Papers 
% that have been accepted for publication should be cited as ``in press'' \cite{b5}. 
% Capitalize only the first word in a paper title, except for proper nouns and 
% element symbols.

% For papers published in translation journals, please give the English 
% citation first, followed by the original foreign-language citation \cite{b6}.

\section*{Acknowledgment}
The Knowledge Foundation supports this work through the OSIR project (reference number 20190081) at Blekinge Institute of Technology, Sweden.

\bibliographystyle{plain}
\bibliography{author.bib}

\begin{thebibliography}{10}

\bibitem{barry2007software}
Evelyn~J Barry, Chris~F Kemerer, and Sandra~A Slaughter.
\newblock How software process automation affects software evolution: a
  longitudinal empirical analysis.
\newblock {\em Journal of Software Maintenance and Evolution: Research and
  Practice}, 19(1):1--31, 2007.

\bibitem{ghadhab2021augmenting}
Lobna Ghadhab, Ilyes Jenhani, Mohamed~Wiem Mkaouer, and Montassar~Ben Messaoud.
\newblock Augmenting commit classification by using fine-grained source code
  changes and a pre-trained deep neural language model.
\newblock {\em Information and Software Technology}, 135:106566, 2021.

\bibitem{gharbi2019classification}
Sirine Gharbi, Mohamed~Wiem Mkaouer, Ilyes Jenhani, and Montassar~Ben Messaoud.
\newblock On the classification of software change messages using multi-label
  active learning.
\newblock In {\em Proceedings of the 34th ACM/SIGAPP Symposium on Applied
  Computing}, pages 1760--1767, 2019.

\bibitem{hindle2009automatic}
Abram Hindle, Daniel~M German, Michael~W Godfrey, and Richard~C Holt.
\newblock Automatic classication of large changes into maintenance categories.
\newblock In {\em 2009 IEEE 17th International Conference on Program
  Comprehension}, pages 30--39. IEEE, 2009.

\bibitem{lehman1980programs}
Meir~M Lehman.
\newblock Programs, life cycles, and laws of software evolution.
\newblock {\em Proceedings of the IEEE}, 68(9):1060--1076, 1980.

\bibitem{lenarduzzi2020open}
Valentina Lenarduzzi, Davide Taibi, Davide Tosi, Luigi Lavazza, and Sandro
  Morasca.
\newblock Open source software evaluation, selection, and adoption: a
  systematic literature review.
\newblock In {\em 2020 46th Euromicro Conference on Software Engineering and
  Advanced Applications (SEAA)}, pages 437--444. IEEE, 2020.

\bibitem{levin2017boosting}
Stanislav Levin and Amiram Yehudai.
\newblock Boosting automatic commit classification into maintenance activities
  by utilizing source code changes.
\newblock In {\em Proceedings of the 13th International Conference on
  Predictive Models and Data Analytics in Software Engineering}, pages 97--106,
  2017.

\bibitem{li2022exploring}
Xiaozhou Li, Sergio Moreschini, Zheying Zhang, and Davide Taibi.
\newblock Exploring factors and metrics to select open source software
  components for integration: An empirical study.
\newblock {\em Journal of Systems and Software}, 188:111255, 2022.

\bibitem{8880574}
Gregorio Robles, Igor Steinmacher, Paul Adams, and Christoph Treude.
\newblock Twenty years of open source software: From skepticism to mainstream.
\newblock {\em IEEE Software}, 36(6):12--15, 2019.

\bibitem{wieringa2014design}
Roel~J Wieringa.
\newblock {\em Design science methodology for information systems and software
  engineering}.
\newblock Springer, 2014.

\bibitem{yilmaz2022quality}
Nebi Y{\i}lmaz and Ay{\c{c}}a Koluk{\i}sa~Tarhan.
\newblock Quality evaluation models or frameworks for open source software: A
  systematic literature review.
\newblock {\em Journal of Software: Evolution and Process}, page e2458, 2022.

\end{thebibliography}
\end{document}